\documentclass[preprint,3p,times,twocolumn]{elsarticle}

\usepackage{color}
\usepackage{epsfig}
\usepackage{amssymb}
\usepackage{amsmath}
\usepackage{dsfont}

\biboptions{sort&compress}

\usepackage{graphicx,hyperref}
\usepackage{bm}
\usepackage{subfigure}

\newcommand{\beq}{\begin{equation}}
\newcommand{\eeq}{\end{equation}}
\newcommand{\nn}{\nonumber \\}

\newcommand\eqn[1]{(\ref{#1})}      
\newcommand\Eqn[1]{Eq.~(\ref{#1})}  

\newcommand{\bK}{{\bf K}}
\newcommand{\bX}{{\bf X}}

\newcommand{\C}{\mathcal{C}}

\newcommand{\bce}{\begin{center}}
\newcommand{\ece}{\end{center}}

\journal{Physics Letters B}

\begin{document}

\begin{frontmatter}



\title{Nonperturbative {\color{black} infrared} enhancement of non-Gaussian correlators\\in de Sitter space}


\author{J. Serreau \corref{cor1}} \ead{serreau@apc.univ-paris7.fr}%
\address{APC, AstroParticule et Cosmologie, Universit\'e Paris Diderot, CNRS/IN2P3, CEA/Irfu, Observatoire de Paris, Sorbonne Paris Cit\'e,\\ 10, rue Alice Domon et L\'eonie Duquet, 75205 Paris Cedex 13, France}

\begin{abstract}
We compute the four-point correlation function of a light $O(N)$ scalar field in de Sitter space in the large-$N$ limit. For superhorizon momentum modes, infrared effects strongly enhance the size of loop contributions. We find that in the deep infrared limit, the latter are of the same order as the tree-level one. {\color{black} The tree-level momentum structure, characteristic of a contact term, gets renormalized by a factor of order unity. In addition loop contributions give rise to a new momentum structure, characteristic of an exchange diagram, corresponding to the exchange of an effective composite scalar degree of freedom.}
\end{abstract}


\end{frontmatter}

\section{Introduction}
\label{sec:intro}

Quantum field theory in curved spaces is a topic of great interest with a long history \cite{BirellDavies}. The case of de Sitter space has attracted a lot of attention both because of its large degree of symmetry and because of its phenomenological relevance for the early inflationary era and for the current accelerated expansion of the universe. Specific phenomena such as gravitational redshift or particle creation imposes one to rethink much of what is known in Minkowski space, starting from the basic notions of particle and vacuum state, even for free fields \cite{Mottola:1984ar}. At present, free gauge fields, such as the photon or the graviton, are still the subjects of debates \cite{Tsamis:2006gj}. 

Interacting fields can be studied by means of perturbation theory \cite{Tsamis:1996qk,Weinberg:2005vy,Sloth:2006az,vanderMeulen:2007ah,Senatore:2009cf,Kahya:2010xh}. They pose practical and conceptual issues. An example is the trans-Planckian problem \cite{Jacobson:1999zk}, i.e., the question of the effective decoupling between infrared and ultraviolet physics, which underlies the very concept of quantum field theory on de Sitter space. They also reveal novel specific features as compared to the flat space case. For instance, scalar fields of sufficiently large mass---in units of the expansion rate---are fundamentally unstable and can decay to themselves \cite{Bros:2006gs}. Light fields, which have no Minkowski analog, are also of great interest because of their phenomenological relevance, e.g., for inflationary cosmology. They exhibit strong semi-classical fluctuations for superhorizon modes and turn out to be essentially nonperturbative, even at weak coupling, due to large infrared effects \cite{Starobinsky:1994bd,Weinberg:2005vy}. In recent years, various methods inspired from flat space techniques have been developed to deal with infrared issues in de Sitter space. Results are still rather scarce but the nonperturbative aspects of light scalar fields are being unravelled \cite{Riotto:2008mv,Garbrecht:2011gu,Burgess:2009bs,Serreau:2011fu,Prokopec:2011ms,Rajaraman:2010xd,Boyanovsky:2012qs,Parentani:2012tx,Serreau:2013psa,Youssef:2013by,Gautier:2013aoa}.

A typical example is the phenomenon of dynamical mass generation: a field with vanishing tree-level mass develops an effective mass due to its self-interactions \cite{Hu:1985uy,Starobinsky:1994bd}. This lifts the flat tree-level potential and regulates possible infrared divergences. Incidentally, this results in nonanalytic coupling dependences of physical observables. A similar phenomenon has been demonstrated for an $O(N)$ scalar field in the large-$N$ limit in the case where the tree-level potential shows spontaneous symmetry breaking \cite{Serreau:2011fu}. Strong infrared fluctuations restore the symmetry, as anticipated in \cite{Ratra:1984yq}, and lead to nonperturbatively enhanced loop contributions \cite{Serreau:2013psa}. 

Immediate phenomenological implications of nontrivial field interactions in the inflationary universe are possible quantum corrections to standard inflationary observables \cite{Boyanovsky:2005px,Sloth:2006az}, or the possibility of non-Gaussian features of primordial density fluctuations \cite{Maldacena:2002vr}. {\color{black} As a first step towards the understanding of the actual cosmological (curvature) perturbations, it often proves useful to consider the simpler case of test scalar fields on a de Sitter background. In this context, it has been pointed out that infrared effects may lead to parametrically enhanced non-Gaussianities at tree-level both for light (massless) fields \cite{Riotto:2008mv} and for the case of a negative tree-level square mass \cite{Serreau:2011fu}. 

The calculation of Ref. \cite{Riotto:2008mv} is based on estimating the four-point correlator of an $O(N)$ scalar field by including loop corrections to the external legs propagators but keeping a simple tree-level interaction vertex. In this Letter, we extend on this and consider loop corrections to the four-point vertex as well. We show that the corresponding contributions to the four-point correlator are also amplified by infrared/secular effects and eventually contribute the same order in coupling as the tree-level contribution. We consider an $O(N)$ theory with quartic self-interactions in the large-$N$ limit. This sums up infinitely many loop diagrams and enables us to capture genuine nonperturbative effects.} Using the expressions for the field propagator and four-point vertex function recently obtained in Refs. \cite{Serreau:2011fu,Serreau:2013psa}, we compute the equal time four-point correlation function for superhorizon modes, which we obtain in closed analytical form. {\color{black} This allows us to analyze the loop contributions in detail and to show that a perturbative treatment fails for superhorizon momenta. We find that radiative corrections give an order one contribution to the tree-level contact term and give rise to an additional momentum structure, characteristic of an exchange diagram.}

\section{General setting}
\label{sec:model}

Consider the $O(N)$-symmetric scalar field theory with classical action (a sum over $a=1,\ldots,N$ is implied)
\beq
\label{eq:action}
 {\cal S}[\varphi]=\int_x\left\{{1\over2}\varphi_a\left(\square-m_{\rm dS}^2\right)\varphi_a-\frac{\lambda}{4!N}\left(\varphi_a\varphi_a\right)^2\right\},
\eeq
with the invariant measure $\int_x\equiv\int d^{d+1}x\,\sqrt{-g}$, on the expanding Poincar\'e patch of a $d+1$-dimensional de Sitter space. In terms of comoving spatial coordinates $\bX$ and conformal time $-\infty<\eta<0$, the line-element reads (we choose the Hubble scale $H=1$)
\beq
 ds^2=\eta^{-2}\left(-d\eta^2+d\bX\cdot d\bX\right).
\eeq
In \Eqn{eq:action}, the mass term $m_{\rm dS}^2=m^2+\xi{\cal R}$ includes a possible coupling to the Ricci scalar ${\cal R}=d(d+1)$ and $\square$ is the appropriate Laplace operator. 

\begin{figure}[t!]  
  \centering
\epsfig{file=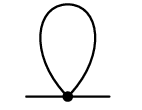,width=2.6cm}
 \caption{\label{fig:Sigma} 
The self-energy in the limit $N\to\infty$; see \Eqn{eq:sigma}. The internal line corresponds to the propagator $G$ itself, hence the nonperturbative character of this limit.}
\end{figure}
 
In the following we consider the $n$-point correlation and vertex functions of the conformally rescaled fields $\phi_a(x)=(-\eta)^{1-d\over2}\varphi_a(x)$ in the (interacting) Bunch Davies vacuum state. The latter are conveniently expressed in terms of time-ordered products of field operators along a closed contour in (conformal) time; see, e.g., \cite{Parentani:2012tx}. For instance the two-point function $G_{ab}(x,x')=\langle T_\C\phi_a(x)\phi_b(x')\rangle$, where $T_\C$ denotes time-ordering along the contour $\C$, encodes both the statistical and spectral correlators $F_{ab}(x,x')={1\over2}\langle\{\phi_a(x),\phi_b(x')\}\rangle$ and $\rho_{ab}(x,x')=i\langle[\phi_a(x),\phi_b(x')]\rangle$:
\beq
\label{eq:Frho}
 G_{ab}(x,x')=F_{ab}(x,x')-\frac{i}{2}{\rm sign}_\C(x^0-x^{\prime0})\rho_{ab}(x,x')\,,
\eeq
where the sign function is to be understood on the contour $\C$. It was shown in \cite{Serreau:2011fu} that, in the large-$N$ limit, the system only admits $O(N)$-symmetric solutions. We thus have $\langle \phi_a\rangle=0$ and $G_{ab}=\delta_{ab} G$. 

In the symmetric phase, the four-point correlation and vertex functions $G^{(4)}$ and $\Gamma^{(4)}$ are related by 
\beq
\label{eq:correlator4}
 G^{(4)}_{ABCD}= G_{AA'}G_{BB'}G_{CC'}G_{DD'}i\Gamma^{(4)}_{A'B'C'D'}
\eeq
where capital letter indices collectively denote space-time variables and $O(N)$ indices and an appropriate integral/summation over repeated indices is understood. Here, we are interested in computing the equal-time four-point correlator in comoving momentum space $G^{(4)}(\eta,\bK_1,\ldots,\bK_4)$ for superhorizon physical momenta, $-K_i\eta\lesssim1$, where $K_i=|{\bf K}_i|$. Both the propagator $G$ and the vertex $\Gamma^{(4)}$ have been computed recently in the infrared regime in the limit $N\to\infty$ \cite{Serreau:2011fu,Serreau:2013psa}. Let us briefly review the results relevant for our present purposes.

In comoving momentum space, the propagator has the free-field-like expression, for ${\rm sign}_\C(\eta-\eta')=1$,
\beq
\label{eq:C4}
 G(\eta,\eta',K)=\frac{\pi}{4}\sqrt{\eta\eta'}H_\nu(-K\eta)H^*_\nu(-K\eta')
\eeq
where $H_\nu(z)$ is the Hankel function of the first kind and $\nu=\sqrt{d^2/4-M^2}$. Here, $M$ a self-consistent, dynamically generated mass, to be discussed shortly. In the cases of interest below, $M\ll1$ and it is convenient to introduce the small parameter $\varepsilon=d/2-\nu\approx M^2/d$. For superhorizon modes, the statistical and spectral two-point function read
\begin{align}
\label{eq:F}
 F_{\rm IR}(\eta,\eta',K)&=\sqrt{\eta\eta'}\frac{F_\nu}{\left(K^2\eta\eta'\right)^{\nu}}\\
\label{eq:rho}
 \rho_{\rm IR}(\eta,\eta',K)&=-\sqrt{\eta\eta'}\,{\cal P}^0_\nu\left(\ln{\eta\over \eta'}\right),
\end{align}
where $F_\nu=[2^\nu\Gamma(\nu)]^2/4\pi$ and we introduced the function
\beq
\label{eq:P}
 {\cal P}^b_a(x)=\frac{\sinh(a x)}{a}e^{-b|x|}.
\eeq

The self-consistent mass $M$ satisfies the gap equation
\beq
\label{eq:gap}
 M^2=m_{\rm dS}^2+\sigma
\eeq
where the constant $\sigma$ is given by the tadpole diagram of Fig. \ref{fig:Sigma}. Retaining only the dominant infrared contribution in the loop (see \cite{Serreau:2011fu} for a complete treatment), one gets
\beq
\label{eq:sigma}
 \sigma=\frac{\lambda}{6N}\langle\varphi^2(x)\rangle\approx\frac{\lambda_{\rm eff}}{\varepsilon},
\eeq
where we introduced $\lambda_{\rm eff}=\lambda F_\nu\Omega_d/12(2\pi)^d$ and $\Omega_d=2\pi^{d/2}/\Gamma(d/2)$. Equation \eqn{eq:gap} is solved as
\beq
\label{eq:gapsol}
 M^2=\frac{m_{\rm dS}^2}{2}+\sqrt{\frac{\left(m_{\rm dS}^2\right)^2}{4}+d\lambda_{\rm eff}}.
\eeq
{\color{black} This produces the known \cite{Sloth:2006az,Starobinsky:1994bd,Garbrecht:2011gu,Burgess:2009bs,Serreau:2011fu} result $M^2\propto \sqrt\lambda$ in the case of light (massless) fields $m_{\rm dS}^2\ll\lambda$.}
The nonanalytic coupling dependence reflects the nonperturbative infrared character of the phenomenon of mass generation. 

\section{Four-point correlator}

The four-point vertex function can be written as \cite{Serreau:2013psa}
\begin{align}
 &\Gamma^{(4)}_{abcd}(\eta_i,\bK_i)=\left[\eta_1\cdots \eta_4\right]^{d-3\over4}\nn
 &\times\Big\{\delta_{ab}\delta_{cd}\,\delta_\C(\eta_1\!-\!\eta_2)\delta_\C(\eta_3\!-\!\eta_4)i D(\eta_1,\eta_3,K_{12})+{\rm perm.}\Big\},\nn
 \label{eq:fourpointconf}
\end{align}
where $\delta_\C(\eta-\eta')$ is a Dirac delta function on the contour, $K_{ij}=|\bK_i+\bK_j|$ and 'perm.' denotes the two permutations needed to make $\Gamma^{(4)}$ symmetric. The function $D$ is the two-point correlator of the composite field $\chi\propto\phi^2$:
\beq
\label{eq:D}
 i D(\eta,\eta',K)=-{\lambda\over3N}\left[\delta_\C(\eta-\eta')+iI(\eta,\eta',K)\right].
\eeq
\begin{figure}[t!]  
\epsfig{file=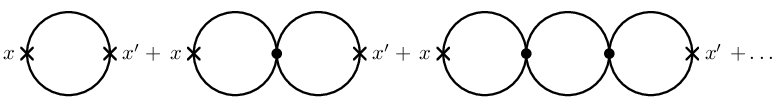,width=7.7cm}
 \caption{\label{fig:bubbles} 
\color{black} The infinite series of multi-bubble diagrams defining the function $I(x,x')\equiv I(\eta,\eta',|\bX-\bX'|)$; see \Eqn{eq:int} in comoving momentum space. The black dots correspond to interaction vertices whereas the crosses denote the endpoints of the function. The one-loop bubble is given by the function $\Pi(x,x')$, see \Eqn{eq:momPi}. Each additional bubble involves a summation of field components and thus comes with a factor $N$, which is compensated by a $1/N$ from the corresponding additional vertex. All such diagrams are thus of the same order in $1/N$ but include arbitrarily high powers of the coupling $\lambda$.}
\end{figure}
The first term on the right hand side corresponds, when inserted in \Eqn{eq:fourpointconf}, to the tree-level vertex and the function $I$ resums an infinite series of bubble loop diagrams, {\color{black} as shown in Fig.~\ref{fig:bubbles}. This resummation is encoded in the following integral equation \cite{Parentani:2012tx}
 \beq
 \label{eq:int}
  I(\eta,\eta',K)=\Pi(\eta,\eta',K)+i\int_\C d\xi\, \,\Pi(\eta,\xi ,K)I(\xi ,\eta',K),
\eeq
where the one-loop contribution $\Pi$ is given by  
\beq
\label{eq:momPi}
 \Pi(\eta,\eta',K)=-\frac{\lambda}{6}\,(\eta\eta')^{d-3\over2}\!\!\int_{\bf Q} G\left(\eta,\eta',Q\right) G\left(\eta,\eta',R\right),
\eeq
with $\int_{\bf Q}=\int d^dQ/(2\pi)^d$ and $R=|{\bf K}+{\bf Q}|$. The function $\Pi$ can be decomposed in a statistical and a spectral components as in \eqn{eq:Frho}. The corresponding momentum integrals in \eqn{eq:momPi} can be evaluated in closed form for infrared physical momenta $|K\eta|,|K\eta'|\lesssim1$ and read \cite{Serreau:2013psa}
\begin{align}
\label{eq:PiF}
 \Pi_F^{\rm IR}(\eta,\eta',K)&=-{\pi_\rho\over\sqrt{\eta\eta'}} \frac{F_\nu}{\left(K^2\eta\eta'\right)^{\kappa}},\\
\label{eq:Pirho}
 \Pi_\rho^{\rm IR}(\eta,\eta',K)&={\pi_\rho\over\sqrt{\eta\eta'}}\,{\cal P}^\varepsilon_{\nu}\left(\ln{\eta\over \eta'}\right),
\end{align}
where $\pi_\rho=2\sigma$ and $\kappa=\nu-\varepsilon$. 

A detailed analysis \cite{Serreau:2013psa} of the integral equation \eqn{eq:int} reveals that, for superhorizon momenta, each additional loop correction to the one-loop result \eqn{eq:PiF}-\eqn{eq:Pirho} is enhanced by large infrared logarithmic contributions which spoil the perturbative expansion. Remarkably, \Eqn{eq:int} can be solved exactly in this regime and these infrared logarithms actually resum to the following modified power laws
\begin{align}
\label{eq:IF}
 I_F^{\rm IR}(\eta,\eta',K)&=-{\pi_\rho\over\sqrt{\eta\eta'}} \frac{F_\nu}{\left(K^2\eta\eta'\right)^{\bar\kappa}},\\
\label{eq:Irho}
 I_\rho^{\rm IR}(\eta,\eta',K)&={\pi_\rho\over\sqrt{\eta\eta'}}\,{\cal P}^\varepsilon_{\bar\nu}\left(\ln{\eta\over \eta'}\right),
\end{align}
with $\bar\nu=\sqrt{\nu^2-\pi_\rho}$ and $\bar\kappa=\bar\nu-\varepsilon$. Clearly, $\pi_\rho$ is the effective parameter which controls the loop expansion. Expanding the above expressions in powers of $\pi_\rho$ generates the whole series of perturbative infrared logarithms. One sees however that the latter breaks down for large time separations $\pi_\rho|\ln \eta/\eta'|\gtrsim1$ and/or deep infrared momenta $\pi_\rho|\ln K^2\eta\eta'|\gtrsim1$. Since the four point vertex \eqn{eq:fourpointconf} is to be involved in time integrals, see \Eqn{eq:correlator4}, which extend all the way from the time where the relevant momenta are superhorizon to the typical time of horizon crossing, it is important to resum these large logarithmic corrections and to employ the resummed functions \eqn{eq:IF}-\eqn{eq:Irho} instead of the perturbative ones \eqn{eq:PiF}-\eqn{eq:Pirho}. In analogy with the parameter $\varepsilon$, we introduce $\bar\varepsilon=d/2-\bar\nu$. In the following, we assume $\varepsilon,\bar\varepsilon,\pi_\rho\ll1$.

Before to embark in the calculation of the four-point correlator \eqn{eq:correlator4}, an important remark is in order. As discussed above, in the deep infrared regime, all loops actually contribute the same order in coupling to the four-point {\it vertex function} \eqn{eq:fourpointconf} -- which is the reason why a nonperturbative approach such as the large-$N$ limit employed here is necessary -- but they are still suppressed by a factor $\pi_\rho$ as compared to the tree-level contribution. However, in contrast to the tree-level vertex, loop terms are nonlocal in time and may lead to enhanced contributions after the relevant time-integrations have been performed in \Eqn{eq:correlator4}. We shall see below that this is indeed what happens and that, for deep infrared modes, the loop contributions to the four-point {\it correlator} are of the same order in coupling as the tree-level contribution.}

\begin{figure}[t!]  
  \centering
\epsfig{file=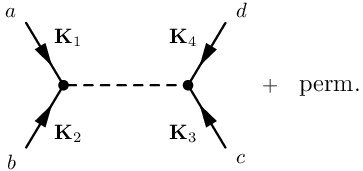,width=5cm}
 \caption{\label{fig:four} 
The equal-time four-point correlator \eqn{eq:G4}. The black lines represent the propagator $G$ and the dashed line represents the nonlocal vertex function \eqn{eq:fourpointconf}, which includes the tree-level vertex and an infinite series of bubble loop diagrams, see Fig. \ref{fig:loops}.}
\end{figure}
 
\begin{figure}[t!]  
  \centering
\epsfig{file=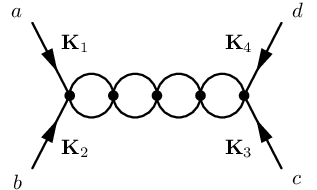,width=4cm}
 \caption{\label{fig:loops} 
A typical multiloop contribution included in the diagram of Fig. \ref{fig:four}. The large-$N$ limit resums the infinite series of such diagrams with an arbitrary number of loops.}
\end{figure}

We now have all the ingredients for our computation of the contribution from superhorizon, infrared modes to the four-point equal-time correlator \eqn{eq:C4}. Writing $\int_\xi=\int_\C d\xi$, the latter can be expressed as the following integral on the time contour $\C$ 
\begin{align}
\label{eq:G4}
 G^{(4)}_{abcd}(\eta,\bK_i)&=\delta_{ab}\delta_{cd}\int_{\xi,\xi'}A_{12}(\eta,\xi)B_{12}(\xi,\xi')A_{34}(\xi',\eta)\nn
  &+{\rm perm.}
\end{align}
where we introduced the functions
\begin{align}
 A_{ij}(\eta,\xi)&=G(\eta,\xi,K_i) G(\eta,\xi,K_j),\\
 B_{ij}(\xi,\xi')&=- \left(\xi\xi'\right)^{d-3\over2}D(\xi,\xi',K_{ij}).
\end{align}
The tree-level contribution is $\propto\delta_\C(\xi-\xi')$ and must be performed separately. It involves the time integral
\begin{align}
 &i\int_\C d\xi \,(-\xi)^{d-3}A_{12}(\eta,\xi)A_{34}(\eta,\xi)\nn
 \label{eq:treelevel}
 &=\int_{-\infty}^\eta \!d\xi \,(-\xi)^{d-3}\left\{A_{12}^F(\eta,\xi)A_{34}^\rho(\eta,\xi)+(12\leftrightarrow34)\right\},
\end{align}
where we introduced the statistical and spectral components of the function $A_{ij}$, as in \eqn{eq:Frho}, and we used standard manipulations on the contour \cite{Berges:2004vw}. Note the symmetry relations $A_{ij}^F(\eta,\xi)=A_{ij}^F(\xi,\eta)$ and $A_{ij}^\rho(\eta,\xi)=-A_{ij}^\rho(\xi,\eta)$. Due to the strong infrared enhancement of the statistical function \eqn{eq:F} as compared to the spectral one \eqn{eq:rho}, we have
\begin{align}
 A_{ij}^F(\eta,\xi)&\approx F(\eta,\xi,K_i) F(\eta,\xi,K_j),\\
 A_{ij}^\rho(\eta,\xi)&=F(\eta,\xi,K_i) \rho(\eta,\xi,K_j)+(i\leftrightarrow j),
\end{align}
where we neglected a term $\propto\rho\rho$ in the first line. This is typical of the classical statistical field regime \cite{Aarts:2001yn,vanderMeulen:2007ah} and reveals, in the present context, the classical stochastic nature of de Sitter infrared fluctuations \cite{Polarski:1995jg}. To estimate the contribution from superhorizon modes we replace the integral $\int_{-\infty}^\eta\to\int_{\eta_0}^\eta$, where $\eta_0$ is such that the relevant momenta are superhorizon:\footnote{\color{black} The parameter $\eta_0$ is thus a (here undetermined) combination of the momenta $K_i$. However, we shall see below that its precise value is of no relevance in the limit of infrared momenta.} $|K_i\eta_0|\lesssim1$. One can then use the expressions \eqn{eq:F} and \eqn{eq:rho} to compute \eqn{eq:treelevel}; see below.

The loop contribution in \eqn{eq:G4} involves the nonlocal function $I$ in \eqn{eq:D}. We write  
\begin{align}
 &\int_\C d\xi d\xi'A_{ij}(\eta,\xi)I(\xi,\xi')A_{kl}(\xi',\eta)\nn
 &=\int_{-\infty}^\eta d\xi \int_{-\infty}^\eta d\xi'A_{ij}^\rho(\eta,\xi)I_F(\xi,\xi')A_{kl}^\rho(\xi',\eta)\nn
 &-\int_{-\infty}^\eta d\xi \int_{\xi}^\eta d\xi'A_{ij}^F(\eta,\xi)I_\rho(\xi,\xi')A_{kl}^\rho(\xi',\eta)\nn
 \label{eq:formula}
 &-\int_{-\infty}^\eta d\xi\int_{\xi}^\eta d\xi'A_{kl}^F(\eta,\xi)I_\rho(\xi,\xi')A_{ij}^\rho(\xi',\eta),
\end{align}
replace again $\int_{-\infty}^\eta\to\int_{\eta_0}^\eta$ and use the infrared behaviors \eqn{eq:IF} and \eqn{eq:Irho}. The calculation is straightforward. 

Extracting a overall factor and introducing the variable $x=\ln(\eta/\eta_0)$, our final result reads
\begin{align}
 G^{(4)}_{abcd}(\eta,\bK_i)&={\lambda\over3N}{F_\nu^3\over2\nu}\frac{(-\eta)^{2-4\nu}(-\eta_0)^{2\varepsilon}}{(K_1\cdots K_4)^{2\nu}}\delta_{ab}\delta_{cd}\,g(x,K_i)\nn
 \label{eq:factor}
 &+{\rm perm.}\,,
\end{align}
with the two momentum structures
\begin{align}
 g(x,K_i)&=g_1(x)\left(K_1^{2\nu}+\cdots+K_4^{2\nu}\right)\nn
 \label{eq:result}
 &+g_2(x)\,\frac{(K_1^{2\nu}+K_2^{2\nu})(K_3^{2\nu}+K_4^{2\nu})}{\left(K_{12}\right)^{2\bar\kappa}}.
\end{align}
The function $g_1(x)$ receives contributions from the tree-level vertex and from the last two lines of \Eqn{eq:formula}, while  $g_2(x)$ is a pure loop contribution coming from the second line of \Eqn{eq:formula}. We find
\begin{align}
\label{eq:g1}
 g_1(x)&={\cal L}_{2\varepsilon}(x)+\frac{\pi_\rho}{2\nu}\frac{{\cal L}_{\varepsilon+\bar\varepsilon}(x)-{\cal L}_{2\varepsilon}(x)}{\bar\varepsilon-\varepsilon},\\
\label{eq:g2}
 g_2(x)&=\frac{\pi_\rho}{2\nu}(-\eta_0)^{2\bar\varepsilon}{\cal L}^2_{\varepsilon+\bar\varepsilon}(x),
\end{align}
where we defined the function ${\cal L}_a(x)=(e^{ax}-1)/a$. The first term on the right hand side of \Eqn{eq:g1} is the tree-level contribution. Loop terms are $\propto\pi_\rho$. 

{\color{black} Here, we kept explicit the exact dependence on $\eta_0$ which comes out of our calculation. As announced, this dependence is suppressed in the limit of small masses and infrared momenta. At leading order in the infrared logarithms, the variable $x=\ln(-f\eta)-\ln(-f\eta_0)\approx \ln(-f\eta)$, where $f$ denotes any combination of the momenta $K_i$ such that $|f\eta_0|\lesssim1$. Clearly the precise form of $f$ is unimportant at leading logarithmic accuracy. Furthermore, the factor $(-\eta_0)^\varepsilon=1+{\cal O}(\varepsilon)$ in \Eqn{eq:factor} and the same is true with the factor $(-\eta_0)^{\bar\varepsilon}$ in \Eqn{eq:g2}. The suppressed dependence on $\eta_0$ is a nontrivial consistency check of our assumption of infrared dominance of the various momentum and time integrals. In the following we systematically neglect ${\cal O}(\varepsilon,\bar\varepsilon)$ corrections unless they are enhanced by large infrared logarithms.}

\section{Discussion}

\begin{figure}[t!]  
  \centering
\epsfig{file=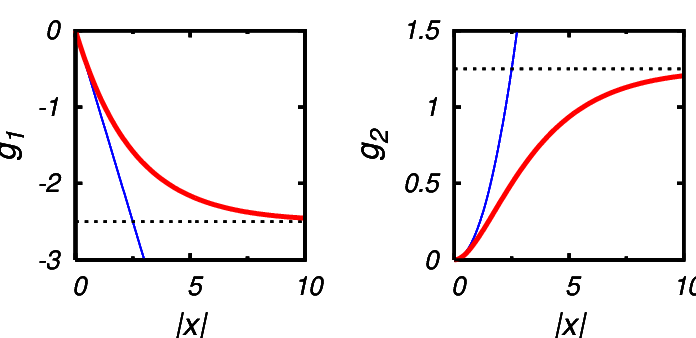,width=7.5cm}
 \caption{\label{fig:g1-g2} \color{black} The functions $g_1$ and $g_2$ in the massless field case with $\varepsilon=0.1$; see \Eqn{eq:g1massless} and \eqn{eq:g2massless}. The light curves show the leading perturbative behaviors for $\varepsilon|x|\ll1$; see \Eqn{eq:pert}. For $\varepsilon|x|\gtrsim1$, loop contributions accumulate and both functions saturate to their respective nonperturbative asymptotic values (dotted lines); see \Eqn{eq:nonpert}.}
\end{figure}

Nontrivial infrared effects arise in the cases of vanishing, or negative tree-level square mass $m_{\rm dS}^2\le0$. For light (massless) fields with $m_{\rm dS}^2\ll\lambda$, the dynamically generated mass \eqn{eq:gapsol} is $M^2\propto\sqrt\lambda$ and
\beq
 \varepsilon=\sqrt{\lambda_{\rm eff}/d}\,,\quad\pi_\rho=2d\varepsilon\,,
\eeq
such that $\bar\varepsilon=3\varepsilon$. {\color{black} As is well-known \cite{Burgess:2009bs,Rajaraman:2010xd,Boyanovsky:2012qs}, loop corrections in that case are controlled by $\pi_\rho\propto\sqrt\lambda$.} We thus get
\begin{align}
\label{eq:g1massless}
 g_1(x)&= {\cal L}_{2\varepsilon}(x)+\varepsilon{\cal L}^2_{2\varepsilon}(x),\\
\label{eq:g2massless}
 g_2(x)&=2\varepsilon{\cal L}^2_{4\varepsilon}(x).
\end{align}
In the regime where infrared logarithms are not too large, $1\lesssim|x|\lesssim 1/\varepsilon$, 
\beq
\label{eq:pert}
 g_1(x)\approx x+\varepsilon x^2\,,\quad g_2(x)\approx2\varepsilon x^2
\eeq
and we {\color{black}precisely recover the usual tree-level result at leading logarithmic accuracy \cite{Bernardeau:2003nx} with $G^{(4)}\sim\lambda x(1+{\cal O}(\varepsilon x))$.} However, for $\varepsilon|x|\sim1$, loop contributions become comparable to the tree-level one and cannot be neglected. In the deep infrared regime, $\varepsilon|x|\gtrsim 1$, the linear growth in $|x|$ saturates and one finds the fully nonperturbative result
\beq
\label{eq:nonpert}
 g_1(x)\approx-{1\over 4\varepsilon}\,,\quad g_2(x)\approx{1\over 8\varepsilon}
\eeq
with $G^{(4)}\sim\sqrt\lambda$. {\color{black} We see that, first, the overall size of the non-Gaussian correlator $G^{(4)}$ is enhanced by a factor $1/\sqrt\lambda$ as compared to the perturbative result \eqn{eq:pert} due to infrared effects\footnote{\color{black} The infrared enhancement of the tree-level four-point correlator had been noticed previously in \cite{Riotto:2008mv} although these authors got a wrong result due to an erroneous manipulations of the limits $|x|\gg1$ and $\varepsilon\ll1$. In our notations, their tree-level result reads $g_1^{\rm RS}(x)=1/2\varepsilon$ and $g_2^{\rm RS}(x)=0$.} and, second, that the contribution from loop diagrams actually contribute the same order in coupling as the tree-level one.\footnote{\color{black} We point out that, although it has been recognized before that, for light fields, the perturbative series is organized in powers of $\sqrt \lambda$ due to infrared effects \cite{Burgess:2009bs,Rajaraman:2010xd,Boyanovsky:2012qs}, here we find that, in the deep infrared regime, tree-level and loop diagrams contribute the same order in the coupling so that there is no perturbative expansion at all.} Indeed, the tree-level contribution alone gives $g_1^{\rm tree}(x)=-1/2\varepsilon$ and $g_2^{\rm tree}(x)=0$. For illustration, the functions \eqn{eq:g1massless} and \eqn{eq:g2massless} are plotted in Fig.~\ref{fig:g1-g2} together with their respective perturbative and nonperturbative limits.}

The other case of interest, where strong infrared effects come into play, is that of spontaneous symmetry breaking at tree-level: $m_{\rm dS}^2<0$. In that case, the symmetry is actually radiatively restored by infrared fluctuations \cite{Serreau:2011fu,Ratra:1984yq}, resulting in a positive effective square mass $M^2\propto\lambda$; see \Eqn{eq:gapsol}. One has, assuming $\lambda\ll|m_{\rm dS}^2|\ll 1$,
\beq
 \varepsilon=\lambda_{\rm eff}/|m_{\rm dS}^2|\,,\quad \pi_\rho=2|m_{\rm dS}^2|.
\eeq
{\color{black} As already pointed out in Refs. \cite{Serreau:2011fu,Serreau:2013psa}, the parameter $\pi_\rho$, which controls the perturbative expansion, is now parametrically of order $\lambda^0$ in the coupling. For\footnote{\color{black}We recall that this is a necessary constraint for the consistency of the present calculation as this guarantees that the result do not depend on the unknown cut-off time $\eta_0$.} $\varepsilon,\bar\varepsilon,\pi_\rho\ll1$, one has $\bar\varepsilon-\varepsilon\approx\pi_\rho/2\nu$ and we get 
\beq
 g_1(x)= {\cal L}_{\bar\varepsilon+\varepsilon}(x)\quad{\rm and}\quad
 g_2(x)=(\bar\varepsilon-\varepsilon){\cal L}^2_{\bar\varepsilon+\varepsilon}(x).
\eeq
We see that in that case loop effects completely dominate the function $g_1$ for any value of $x$. In the deep infrared regime $(\bar\varepsilon+\varepsilon)|x|\gtrsim1$, one has the fully nonperturbative result
\beq
 g_1(x)\approx -{1\over {\bar\varepsilon+\varepsilon}}\,,\quad g_2(x)\approx{\bar\varepsilon-\varepsilon\over(\bar\varepsilon+\varepsilon)^2},
\eeq
which exhibits, again, infrared enhancement: $G^{(4)}\sim\lambda/(\bar\varepsilon+\varepsilon)$.}

Finally, we note the specific momentum dependence of the $g_2$ loop contributions in \eqn{eq:factor}, \eqn{eq:result} which is singular whenever the sum of any two momenta approaches zero, $K_{ij}\to0$. This loop contribution thus gives a distinct signature from the tree-level one and, in fact, provides the dominant contribution for such momentum configurations. This is a direct consequence of the infrared behavior \eqn{eq:IF} of the nonlocal four-point vertex \eqn{eq:fourpointconf}. 
In the present case, the latter is given by the two-point function \eqn{eq:D} of the operator $\phi^2$ which, at large momentum separation is essentially that of a free scalar field of mass $\bar M^2\approx d(\varepsilon+\bar\varepsilon)$, as noticed in \cite{Serreau:2013psa}. {\color{black} 
The $g_2$ loop contribution can thus be seen as describing the exchange of a light (composite) scalar degree of freedom, whereas the $g_1$ term, which receives contribution from both the tree-level vertex and loop corrections, is a contact term \cite{Bernardeau:2003nx}. 

In conclusion, we have obtained an analytic expression of the non-Gaussian four-point correlator of an $O(N)$ scalar field in the large-$N$ limit. Loop contributions get dramatically amplified by infrared/secular effects and, for deep superhorizon momenta, eventually contribute the same order in coupling as the tree-level vertex. The present $O(N)$ scalar field theory in the large-$N$ limit provides an example where such infrared/secular effects can be explicitly resummed, demonstrating how the secular $\ln\eta$ growth of perturbative contributions eventually saturate to well-defined, albeit nonperturbative expressions. We believe our results add to the understanding of the nontrivial infrared physics of light scalar fields in de Sitter space. On the phenomenological side, although it should be made clear that the present test scalar field setup is, by no means, a realistic model of actual cosmological curvature perturbation,\footnote{\color{black} For instance, the relevant self-interactions of curvature perturbations in single field inflation models have both cubic and quartic contributions, with derivative couplings which ameliorate the infrared/late-time behavior of momentum/time integrals in loop diagrams \cite{Maldacena:2002vr,Arroja:2008ga}.} our calculation illustrates how infrared effects can spoil the usual perturbative expectations (see also \cite{Cogollo:2008bi}). Possible implications, e.g., for multifield inflationary models \cite{Bartolo:2001cw}, remain to be investigated.}

\section*{Acknowledgements}
We acknowledge interesting discussions with F. Gautier and R. Parentani.






\end{document}